\documentclass[aps,prl,twocolumn,times,amsmath,amssymb,superscriptaddress,floatfix]{revtex4}

\usepackage{times}
\usepackage{float}
\usepackage{amsmath}
\usepackage{amsthm}
\usepackage{amssymb}
\usepackage{amsbsy}
\usepackage{graphicx}
\usepackage{pstricks}
\usepackage{color}
\usepackage{cancel}
\usepackage{subfigure} 
\usepackage{enumerate}
\usepackage{mathrsfs} 
\usepackage{multirow} 
\usepackage{ulem}
\usepackage{wasysym}
\usepackage{sidecap}

\begin{document}

\title{The quantum origins of moment fragmentation in Nd$_2$Zr$_2$O$_7$}

\author{Owen Benton}
%
\affiliation{Okinawa Institute of Science and Technology Graduate University, Onna-son,
Okinawa 904-0495, Japan}

\begin{abstract}
Spin liquid states are often described as the antithesis of 
magnetic order.
Recently, however, it has been proposed that in certain frustrated magnets the magnetic
degrees of freedom may ``fragment'' in such a way as to give rise to a coexistence of spin liquid
and ordered phases.
Recent neutron scattering results 
[S. Petit et al., Nature Phys., Advance online publication, (2016)]
suggest that this scenario may be realized in the pyrochlore 
magnet Nd$_2$Zr$_2$O$_7$.
These observations show
the characeristic pinch point features of a Coulombic
spin liquid occurring 
alongside the Bragg peaks of an ``all-in-all-out'' 
ordered state.
Here we explain the quantum origins of this apparent 
magnetic moment fragmentation, within the framework
of a quantum model of nearest neighbour exchange, appropriate to Nd$_2$Zr$_2$O$_7$.
This model is able to capture both the ground state  
order and the pinch points observed at finite energy.
The observed fragmentation arises due to the combination of 
the unusual symmetry properties of 
the Nd$^{3+}$ ionic wavefunctions and 
the structure of equations of motion 
of the magnetic degrees of freedom.
The results of our analysis suggest that Nd$_2$Zr$_2$O$_7$ is 
proximate to a $U(1)$ spin liquid phase and is a promising candidate for the
observation of a Higgs transition in a magnetic system.
\end{abstract}

\maketitle

The study of frustrated magnets has uncovered many experimental
systems in which conventional magnetic order is suppressed--
or avoided entirely-- opening the door to novel quantum
ground states \cite{greedan01, gardner10,shimizu03, han12, kimura13, sibille15}.
A particularly enticing possibility is to realize a spin-liquid ground state
possessing emergent gauge fields
and excitations with fractional quantum numbers \cite{balents10}.
As such, spin liquids
provide beautiful examples of the emergence of
new and unexpected degrees of freedom, out of the collective
behaviour of a strongly interacting \mbox{system~\cite{wen02, kitaev06, 
castelnovo08, moessner10, henley10, yan11, gingras14, benton16}.}

While a spin liquid ground state is usually discussed as
an alternative to magnetic order, it has been proposed that 
spin liquid physics can coexist with a conventional magnetic
order parameter \cite{savary12, brooks14, powell15, jaubert15, canals16, paddison-arXiv}.
One way that this can occur \cite{brooks14} is through a ``fragmentation''
of the magnetisation field into two quasi-independent sets of degrees
of freedom, one of which orders and the other of which remains
fluctuating in a spin-liquid-like manner.
In a spin-ice system, such as that considered in Ref. \cite{brooks14},
such a state would be revealed in neutron scattering
experiments by the coexistence of magnetic Bragg peaks, indicating long range
order, with pinch-point singularities.
%
Pinch-points in neutron scattering measurments are known to be
the characteristic of
a Coulomb phase \cite{isakov04, henley05, fennell09}, which hosts
an emergent ``magnetic flux'' obeying Gauss' law and an associated
``electromagnetic'' gauge field.

%

\begin{figure}
\centering
\subfigure[ ]{
\includegraphics[width=0.4\columnwidth]{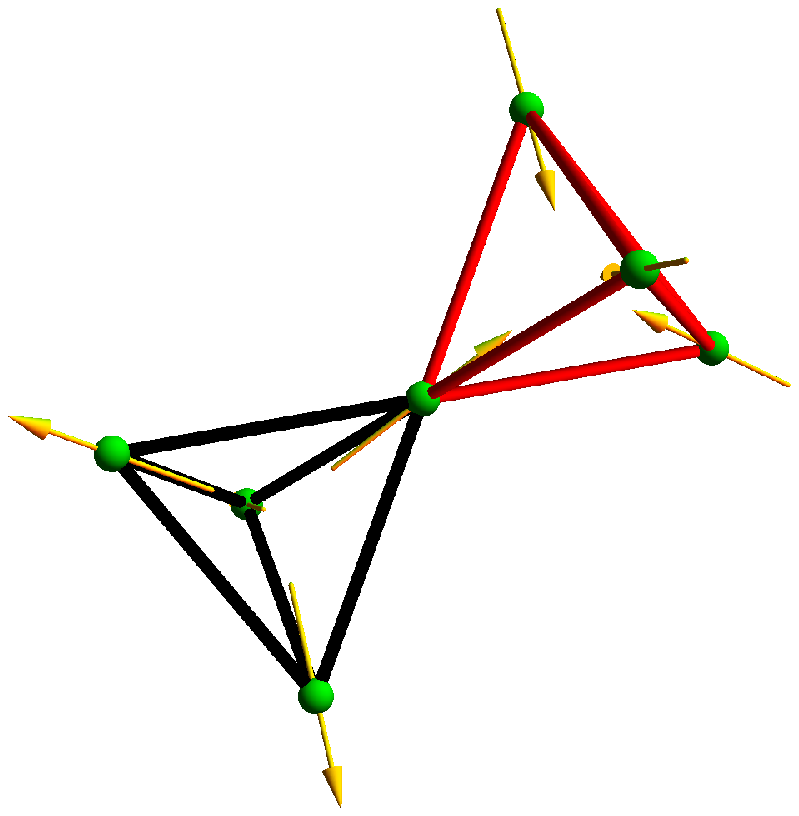}} \
\subfigure[ ]{
\includegraphics[width=0.4\columnwidth]{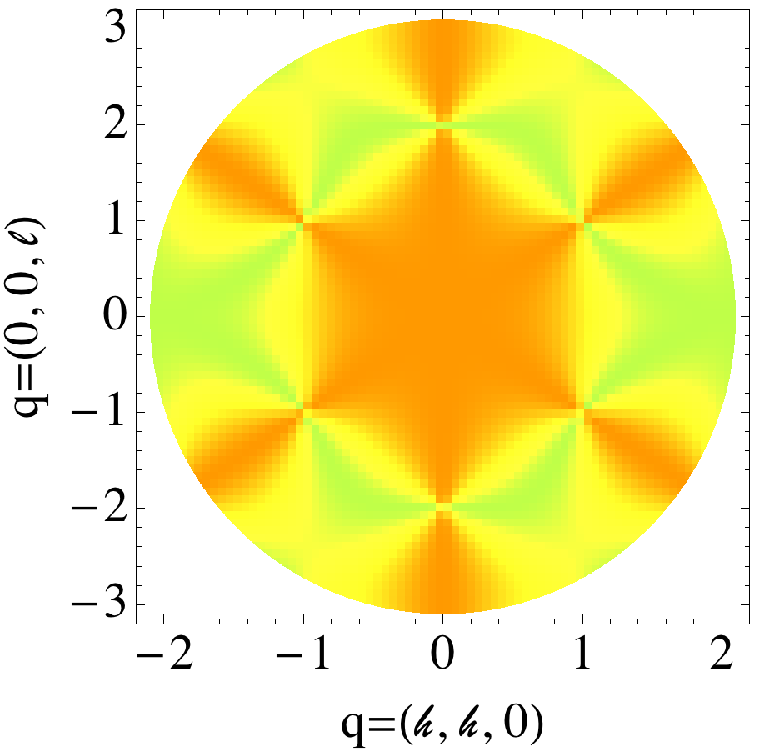} \
\includegraphics[width=0.1\columnwidth]{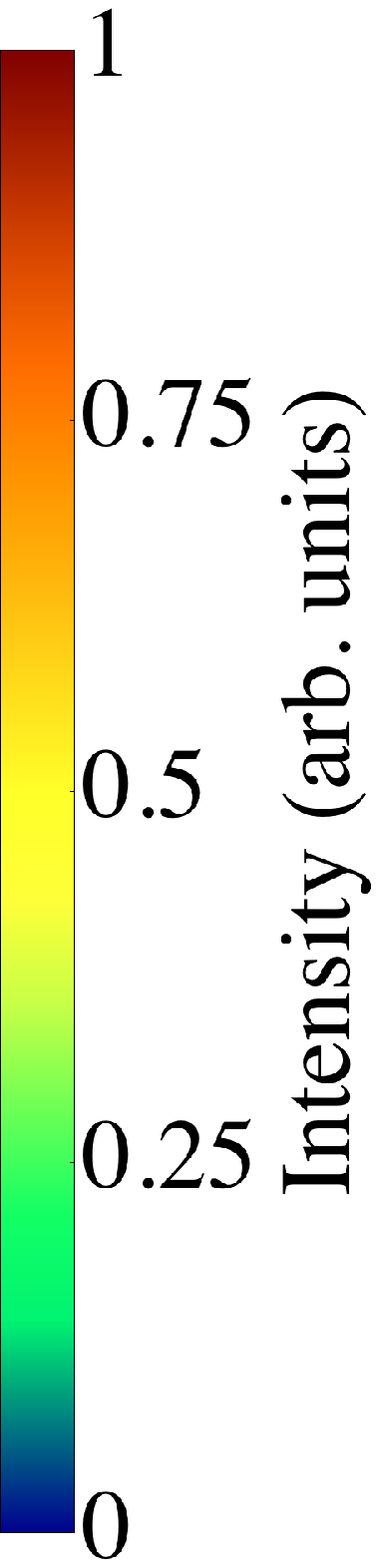}}
\caption{
(a) 
All-in-all-out configuration of magnetic moments on the pyrochlore lattice.
The blue dashed line indicates the local z-axis on the central site.
(b)
Pinch point singularities in the neutron scattering
structure factor 
$\mathcal{S}(\mathbf{q}, \omega=\Delta_{\sf flat})$ [Eq. \ref{eq:Sq-def}]
at finite energy 
above the all-in-all-out ground state in dipolar-octupolar magnets.
These pinch points are reminiscent of those predicted in the ``Coulomb phase''
of spin ice \cite{isakov04, henley05, fennell09}.
Pinch points of this form were observed at energy $\Delta\approx0.07$meV
in recent experiments on the pyrochlore material Nd$_2$Zr$_2$O$_7$ \cite{petit16},
along with Bragg peaks signifying an all-in-all-out ordered phase.
This suggests that  Nd$_2$Zr$_2$O$_7$ exhibits the phenomenon of ``moment fragmentation'',
proposed in Ref. \cite{brooks14}.
Calculation of the structure factor in the flat band was performed using a linear spin wave treatment 
of the exchange Hamiltonian $\mathcal{H}_{\sf XYZ}^{\sf DO}$ [Eq. (\ref{eq:HXYZ})] and 
the exchange parameters in Eq. (\ref{eq:Jparams}).
}
\label{fig:pinchpoints}
\end{figure}

In a 
remarkable experimental development 
the signatures of this fragmentation 
have recently been observed in the
pyrochlore magnet Nd$_2$Zr$_2$O$_7$ \cite{petit16}.
Nd$_2$Zr$_2$O$_7$ is known to undergo magnetic 
ordering at a temperature $T_N\approx0.3$K \cite{bloete69, xu15, lhotel15},
with the formation of magnetic Bragg peaks consistent with
an antiferromagnetic, ``all-in-all-out'', ordered state shown in Fig. \ref{fig:pinchpoints}(a).
Alongside these Bragg peaks, the authors of Ref.~\cite{petit16}
observed pinch point singularities, like those shown
in Fig. \ref{fig:pinchpoints}(b).
These pinch points occur as part of a flat band at finite 
energy $\Delta \approx 0.07$ meV.

The observation of pinch points, signifying the physics of a Coulomb
phase, against the background of these Bragg peaks
would seem to provide compelling evidence for the realization of the
fragmentation scenario proposed in Ref. \cite{brooks14}.
While calculations using the random phase approximation (RPA)
presented in Ref. \cite{petit16} were able to capture the pinch points,
the parameters of the fitted model were not compatible with an all-in-all-out
ground state.
More broadly than this,  the 
question of the mechanism of the moment
fragmentation
 in Nd$_2$Zr$_2$O$_7$ remains open.

Here, we explain the quantum origins of the moment fragmentation
observed in Nd$_2$Zr$_2$O$_7$.
This fragmentation
 is the combined consequence of the
``dipolar-octupolar''~\cite{huang14}  nature of the Nd$^{3+}$ 
Kramers doublets and of the structure of the equations of motion for the
pseudospin operators describing those doublets.
Our theory goes beyond previous work by reconciling
the magnetic ground state of
Nd$_2$Zr$_2$O$_7$ with its observed spectrum in inelastic neutron
scattering, including the presence of pinch points.
Through this theory we are able to reveal the true nature of the 
moment fragmentation and find that Nd$_2$Zr$_2$O$_7$ is
proximate to a $U(1)$ spin liquid phase.

{\it Model--} To  model the magnetism of Nd$_2$Zr$_2$O$_7$
we must begin with the physics of Nd$^{3+}$ ions in their
local crystal field. 
The ground state of the crystal field is a Kramers doublet, separated
by a gap of $\Delta_{\sf CF} \approx 23$ meV from the lowest excited
doublet \cite{xu15}.
This doublet has dipolar-octupolar character~\cite{huang14, xu15, lhotel15,
ciomaga15}, a fact which will have important consequences in our discussion.

A natural choice of basis $\{|\uparrow_z\rangle, |\downarrow_z\rangle\}$ 
for this doublet is the one which diagonalizes
the $z$-component of the angular momentum operator ${\bf J}$, where
the $z$-axis is defined locally as the $C_3$ symmetry axis pointing 
from a magnetic site through the centres of the two pyrochlore
tetrahedra which share it [cf. Fig. \ref{fig:pinchpoints}(a)].
While $J_z$ has finite matrix elements within the doublet, the planar
components $\{ J_x, J_y \}$ both vanish \cite{xu15, lhotel15}.

To describe the interactions of these dipolar-octupolar doublets
we introduce on each site $i$ a 
vector of  pseudospin-1/2 operators $\vec{\tau}_i=(\tau^x_i, \tau^y_i,
\tau^z_i)$.
Since $\langle J_x \rangle =\langle J_y \rangle=0$ within the doublet,
the magnetisation on site $i$ is given by
\begin{eqnarray}
{\bf m}_i= g_z \mu_B \tau_i^z \hat{{\bf z}}_i
\label{eq:magnetisation-1}
\end{eqnarray}
where $\hat{{\bf z}}_i$ is a unit vector in the local
$z$-direction and $g_z$ is the $z$-component of the g-tensor.

As discussed in Ref. \cite{huang14},
the symmetry properties of $\vec{\tau}$ are somewhat counter-intuitive:
both $\tau^z_i$ and $\tau^x_i$ transform like the $z$-component of a magnetic
dipole moment. 
The operator $\tau^y_i$ meanwhile, transforms like an element of the magnetic 
octupole tensor.
From these symmetry properties one can deduce the most general form of 
nearest-neighbour interactions between the operators 
$\tau^{\alpha}_i$ allowed by the symmetries of
the system \cite{huang14, carrasquilla15} (time reversal $\otimes$ lattice symmetries):
\begin{eqnarray}
&&\mathcal{H}_{\sf ex}^{\sf DO}=
\sum_{\langle ij \rangle}
\bigg[
{\sf J}_x \tau^x_i \tau^x_j
+
{\sf J}_y \tau^y_i \tau^y_j
+
{\sf J}_z \tau^z_i \tau^z_j
\nonumber \\
&&\qquad\qquad
+
{\sf J}_{xz}
\left( 
 \tau^x_i \tau^z_j
+
 \tau^z_i \tau^x_j
\right)
\bigg]
\label{eq:H1}
\end{eqnarray}

The interaction ${\sf J}_{xz}$ may be removed by
a global pseudospin rotation $\tau^{\alpha}_i\to \tilde{\tau}^{\tilde{\alpha}}_i$
where 
\begin{eqnarray}
&&\tilde{\tau}^{\tilde{x}}_i=\cos(\vartheta) \tau^x_i+\sin(\vartheta) \tau^z_i; \ 
\tilde{\tau}^{\tilde{y}}_i=\tau^y_i; \ 
\nonumber \\
&&
\tilde{\tau}^{\tilde{z}}_i=\cos(\vartheta) \tau^z_i-\sin(\vartheta) \tau^x_i; \
 \tan(2\vartheta)= \frac{2 {\sf J}_{xy}}{{\sf J}_x-{\sf J}_z}.
\label{eq:theta}
\end{eqnarray}
This leaves us with an ``$XYZ$'' Hamiltonian for the rotated pseudospins $\tilde{\tau}^{\tilde{\alpha}}$
\begin{eqnarray}
\mathcal{H}_{\sf XYZ}^{\sf DO}=
\sum_{\langle ij \rangle}
\bigg[
\tilde{\sf J}_x \tilde{\tau}^{\tilde{x}}_i \tilde{\tau}^{\tilde{x}}_j
+
\tilde{\sf J}_y \tilde{\tau}^{\tilde{y}}_i \tilde{\tau}^{\tilde{y}}_j
+
\tilde{\sf J}_z \tilde{\tau}^{\tilde{z}}_i \tilde{\tau}^{\tilde{z}}_j
\bigg].
\label{eq:HXYZ}
\end{eqnarray}
The phase diagram of $\mathcal{H}_{\sf XYZ}^{\sf DO}$ is then
a function of the three parameters $\tilde{\sf J}_{\tilde{x}, \tilde{y}, \tilde{z}}$
and does not depend on the angle $\vartheta$.

This does not mean, however, that $\vartheta$ plays no further role in
the physics of the system.
The magnetisation on each site, in terms of the rotated 
pseudospins $\tilde{\tau}^{\tilde{\alpha}}_i$
is
\begin{eqnarray}
{\bf m}_i= g_z \mu_B (\cos(\vartheta)\tilde{\tau}_i^{\tilde{z}} + 
\sin(\vartheta)\tilde{\tau}_i^{\tilde{x}}) \hat{{\bf z}}_i.
\label{eq:magnetisation-2}
\end{eqnarray}
The angle $\vartheta$ thus controls how the 
pseudospins $\tilde{\tau}^{\tilde{\alpha}}_i$ couple to an 
external probe which scatters off the internal magnetic fields--
such as a neutron. 

The peculiar moment fragmentation observed in Nd$_2$Zr$_2$O$_7$ stems
from Eq. (\ref{eq:magnetisation-2}).
Eq. (\ref{eq:magnetisation-2}) can be split up in terms of its contributions from
 $\tilde{\tau}^{\tilde{z}}_i$ and  $\tilde{\tau}^{\tilde{x}}_i$ 
\begin{eqnarray}
&&{\bf m}_i= g_z \mu_B \cos(\vartheta) {\bf m}_i^{(\tilde{z})}
+g_z \mu_B \sin(\vartheta) {\bf m}_i^{(\tilde{x})}  
\label{eq:magnetisation-3}
\\
&&
{\bf m}_i^{(\tilde{\alpha})}=
\tilde{\tau}_i^{\tilde{\alpha}} \hat{{\bf z}}_i
. \label{eq:xyz-mfields}
\end{eqnarray}
In essence, the origin of the fragmentation is that 
${\bf m}_i^{(\tilde{z})}$  orders, forming the
all-in-all-out order which is responsible for the observed magnetic
Bragg peaks, while fluctuations of ${\bf m}_i^{(\tilde{x})}$
get shifted to finite energy.
These fluctuations of ${\bf m}_i^{(\tilde{x})}$
themselves decouple dynamically into a flat band obeying 
$\nabla \cdot {\bf m}_i^{(\tilde{x})}=0$ and therefore
exhibiting the correlations of a Coulomb phase, and two higher energy 
dispersive bands.

\begin{figure*}
\subfigure[\ \  $\mathcal{S}({\bf q}=(h, h, 2), \omega) $]{
\includegraphics[width=0.4\columnwidth]{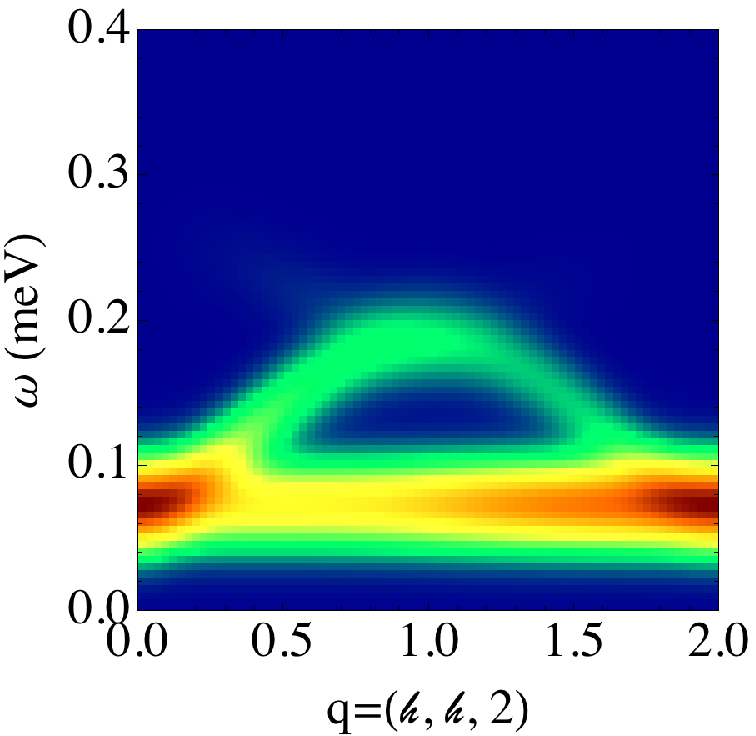}} \
\subfigure[\ \  $\mathcal{S}({\bf q}=(h, h, 0), \omega)$ ]{
\includegraphics[width=0.4\columnwidth]{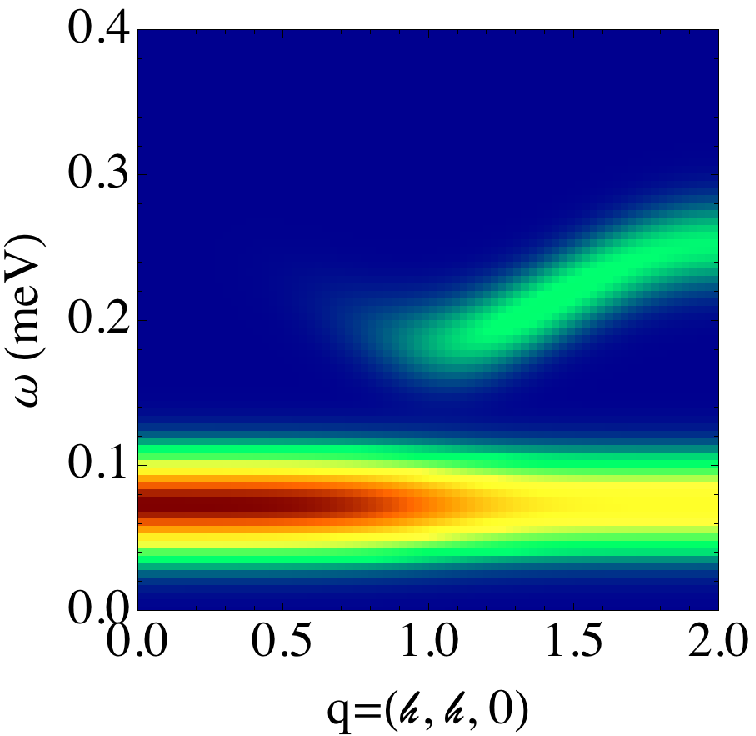}} \
\subfigure[\ \  $\mathcal{S}({\bf q}=(h, h, 2-h), \omega)$ ]{
\includegraphics[width=0.4\columnwidth]{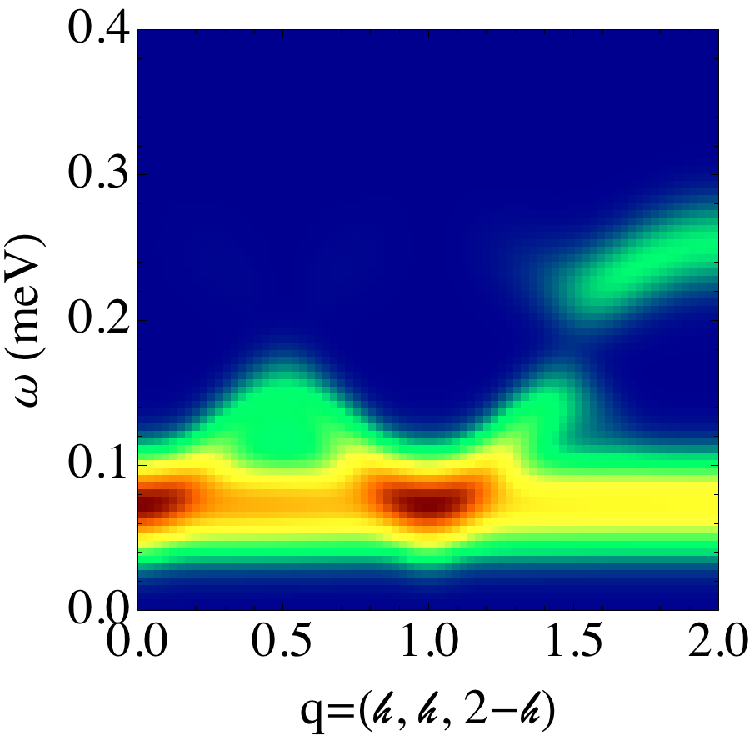}} \
\subfigure[\ \  $ \mathcal{S}({\bf q}=(1, 1, l), \omega)$ ]{
\includegraphics[width=0.4\columnwidth]{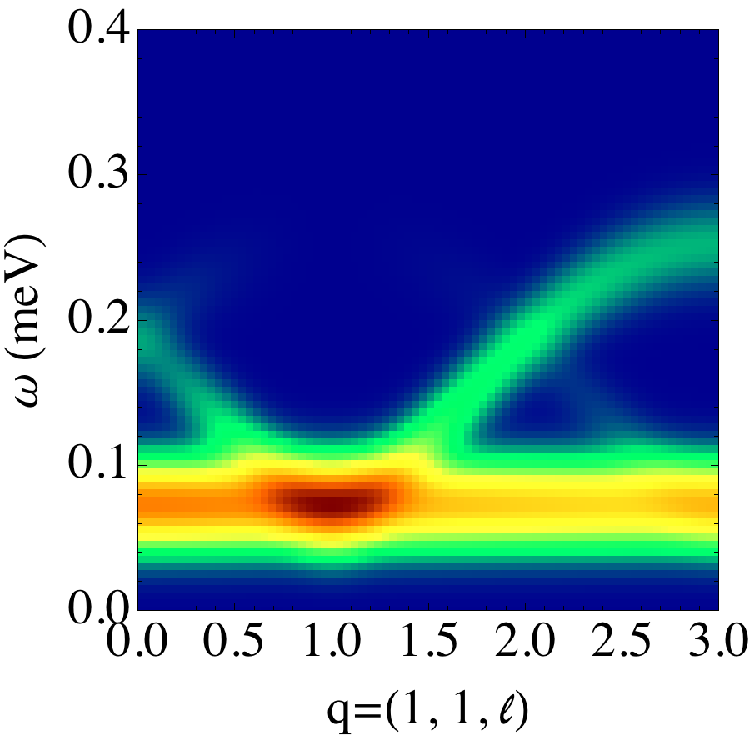}} \
\includegraphics[width=0.1\columnwidth]{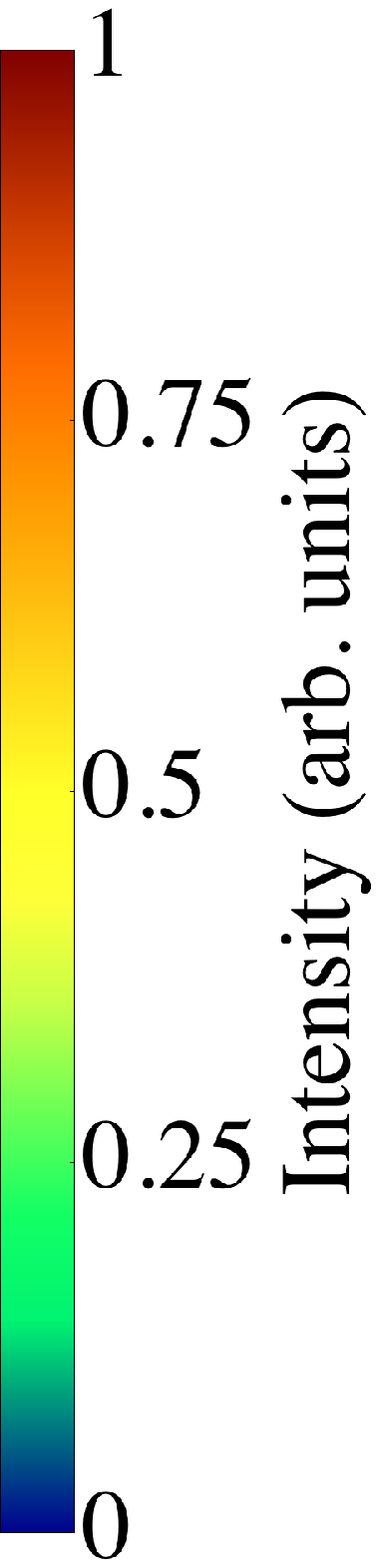} 
\caption{Structure factor for inelastic neutron scattering [Eq. (\ref{eq:Sq-def})] in Nd$_2$Zr$_2$O$_7$
calculated from linear spin wave theory. 
Calculations are made from a
linear spin wave treatment of $\mathcal{H}_{\sf XYZ}^{\sf DO}$ [Eq. (\ref{eq:HXYZ})]
, assuming an
all-in-all-out ground state order 
and with exchange parameters given in Eq. (\ref{eq:Jparams}).
The flat band at $\Delta_{\sf flat}\approx0.07$meV contains the pinch point structures shown
in Fig. \ref{fig:pinchpoints}(b).
The calculated spin wave dispersion and intensities are in good agreement with the inelastic
scattering measurements in Ref. \cite{petit16}.
The structure factor was calculated at
$T=0$ and convoluted with a Gaussian of
full width at half-maximum $=0.05$ meV to mimic finite experimental resolution.
}
\label{fig:inelastic}
\end{figure*}

{\it Spin wave theory}--
To see this we will first use a spin wave expansion around the
all-in-all-out ground state.
An all-in-all-out ground state, with
$\langle \tilde{\tau}_i^{\tilde{z}}  \rangle \neq 0$ and 
$\langle \tilde{\tau}_i^{\tilde{x}} \rangle =
\langle \tilde{\tau}_i^{\tilde{y}} \rangle = 0$
is a classical ground state of
$\mathcal{H}_{\sf XYZ}^{\sf DO}$
 [Eq. (\ref{eq:HXYZ})] when
\begin{eqnarray}
\tilde{\sf J}_z<0, \quad
-|\tilde{\sf J}_z|<\tilde{\sf J}_{x}, \tilde{\sf J}_y<3 |\tilde{\sf J}_z|.
\end{eqnarray}
The spin wave expansion proceeds by introducing Holstein-Primakoff
bosons $[a_i, a_j^{\dagger}]=\delta_{ij}$
and writing
\begin{eqnarray}
&&\tilde{\tau}^{\tilde{z}}_i=S-a_i^{\dagger}a_i  
\label{eq:HPtauz}
\\
&&\tilde{\tau}^{+}_i
=\tilde{\tau}^{\tilde{x}}_i+i \tilde{\tau}^{\tilde{y}}_i=\sqrt{\left( 2S-a_i^{\dagger}a_i \right)} \ \ a_i
\label{eq:HPtau+}
\approx \sqrt{2S} a_i \\
&&\tilde{\tau}^{-}_i
=\tilde{\tau}^{\tilde{x}}_i-i \tilde{\tau}^{\tilde{y}}_i= a_i^{\dagger} \sqrt{\left( 2S-a_i^{\dagger}a_i \right)}
\approx \sqrt{2S} a_i^{\dagger}.
\label{eq:HPtau-}
\end{eqnarray}
where $S=\frac{1}{2}$ since we are dealing with pseudospin-$\frac{1}{2}$ operators.

Inserting Eqs.~(\ref{eq:HPtauz})-(\ref{eq:HPtau-}) into Eq.~(\ref{eq:HXYZ}) and keeping
terms only up to bilinear order in $a_i, a_i^{\dagger}$ yields the linear spin wave 
Hamiltonian:
\begin{eqnarray}
&&\mathcal{H}^{\sf DO}_{\sf LSW}=-3 N  |\tilde{J}_z| S^2+
6 |\tilde{\sf J}_z| S
\sum_{i} a_{i}^{\dagger} a_i \nonumber \\
&& \quad
+  \frac{S}{2}
 \sum_{\langle ij \rangle}
(a_i^{\dagger}, \ a_i) 
\begin{pmatrix}
\tilde{{\sf J}}_x + \tilde{{\sf J}}_y & \tilde{{\sf J}}_x-\tilde{{\sf J}}_y\\
 \tilde{{\sf J}}_x-\tilde{{\sf J}}_y & \tilde{{\sf J}}_x + \tilde{{\sf J}}_y 
\end{pmatrix}
\begin{pmatrix}
a_j \\
 a_j^{\dagger}
\end{pmatrix}
\label{eq:HLSW}
\end{eqnarray}
After Fourier transformation into momentum (${\bf q}$) space
 and
a subsequent Bogoliubov transformation \cite{fazekas99, roger83} 
to a new
set of bosonic operators $[b_{\lambda\mathbf{q}}, b_{\lambda'\mathbf{q}'}^\dagger]=\delta_{\mathbf{q} \mathbf{q}'}\delta_{\lambda\lambda'}$
we arrive at a diagonalized Hamiltonian for four bands $\lambda$ of bosons
with dispersion $\omega_{\lambda}(\mathbf{q})$:
\begin{eqnarray}
&&\mathcal{H}^{\sf DO}_{\sf LSW}=-3N |\tilde{J}_z| S(S+1)+\nonumber \\
&& \qquad\qquad\qquad \sum_{\mathbf{q}, \lambda} 
\omega_{\lambda}(\mathbf{q})
\left( b_{\mathbf{q} \lambda}^{\dagger} b^{\phantom\dagger}_{\mathbf{q} \lambda}+
\frac{1}{2}\right).
\label{eq:HLSW-diag}
\end{eqnarray}
These four bands $\omega_{\lambda}(\mathbf{q})$ consist of two degenerate flat bands at energy $\Delta_{\sf flat}=\sqrt{(3|\tilde{\sf J}_z|-\tilde{\sf J}_x)(3|\tilde{\sf J}_z|-\tilde{\sf J}_y)}$
and two dispersive bands.

We can invert the transformations used in obtaining Eq.~(\ref{eq:HLSW-diag}) to calculate the
dynamical correlations accessible in neutron scattering, in terms of the expectation values of
bosonic bilinears.
The structure factor for magnetic neutron scattering is 
\begin{eqnarray}
&&\mathcal{S}(\mathbf{q}, \omega)=\int dt  \ e^{-i \omega t} \sum_{\mu \nu}
\left(
\delta_{\mu \nu}-\frac{q_{\mu} q_{\nu}}{q^2}
\right) \nonumber \\
&& \qquad \qquad \times
\langle m^{\mu}(-\mathbf{q}, 0) m^{\nu}(\mathbf{q}, t)
\rangle
\label{eq:Sq-def}
\end{eqnarray}
where ${\bf m}(\mathbf{q}, t)$ is the lattice Fourier transform
of the site magnetisation [Eq. (\ref{eq:magnetisation-2})]
at time $t$.

The structure factor $S(\mathbf{q}, \omega)$
breaks up into two contributions 
\begin{eqnarray}
\mathcal{S}(\mathbf{q}, \omega)=\mathcal{S}^{(\tilde{z})}(\mathbf{q}, \omega)
+\mathcal{S}^{(\tilde{x})}(\mathbf{q}, \omega).
\label{eq:Sq-z+x}
\end{eqnarray}
The first contribution $\mathcal{S}^{(\tilde{z})}(\mathbf{q}, \omega)$
comes from the ordered and static
correlations of ${\bf m}_i^{(\tilde{z})}$ and gives rise
to magnetic Bragg peaks at $\omega=0$.

The second contribution to $\mathcal{S}(\mathbf{q}, \omega)$ in Eq. (\ref{eq:Sq-z+x})
comes from the dynamic correlations of ${\bf m}_i^{(\tilde{x})}$,
which have the form
\begin{eqnarray}
&&\mathcal{S}^{(\tilde{x})}(\mathbf{q}, \omega)\approx
\sin^2(\vartheta) g_z^2 \mu_B^2 
\sum_{\lambda}
s_{\lambda}(\mathbf{q})
\nonumber \\
&&
\big[ (1+n_B(\omega))\delta(\omega-\omega_{\lambda}(\mathbf{q}))
+n_B(\omega)\delta(\omega+\omega_{\lambda}(\mathbf{q})) \big] \nonumber \\
\label{eq:Sq-taux}
\end{eqnarray}
where $s_{\lambda}(\mathbf{q})$ are coefficients calculated from the
Bogoliubov transformation.

Due to the flat bands,
$\mathcal{S}^{(\tilde{x})}(\mathbf{q}, \omega)$ exhibits a peak at
\mbox{$\omega=\Delta_{\sf flat}$} for all $\mathbf{q}$.
%
The intensity of this flat band peak is plotted in Fig. \ref{fig:pinchpoints}(b).
For all choices of exchange parameters $\tilde{\sf J}_{\alpha}$
within the all-in-all-out phase
it exhibits a pattern of pinch-point singularities, as observed 
at finite energy in Nd$_2$Zr$_2$O$_7$ \cite{petit16}.
Our calculations thus simultaneously reproduce the observation
of pinch points and the correct ground state order for Nd$_2$Zr$_2$O$_7$,
something which was not done previously.

The values of the exchange parameters $\tilde{\sf J}_{\alpha}$ can be
constrained by considering the  inelastic spectrum, which was
measured in~Ref.~\cite{petit16}.
The model used in Ref.~\cite{petit16}
is equivalent to taking Eq. (\ref{eq:HXYZ}) with parameters
\begin{eqnarray}
 \tilde{\sf J}_x=0; \
 \tilde{\sf J}_y=-0.047 \ \text{meV}; \
 \tilde{\sf J}_z=0.103 \  \text{meV}.
\label{eq:Jparams-Petit} 
\end{eqnarray}
This parameterisation 
gives a good description of the
inelastic spectrum but incorrectly predicts an octupolar ground state.
Here, we take instead
\begin{eqnarray}
 \tilde{\sf J}_x=0.103 \  \text{meV}; \
 \tilde{\sf J}_y=0;  \
 \tilde{\sf J}_z=-0.047 \ \text{meV}.
\label{eq:Jparams} 
\end{eqnarray}
This transformed set of parameters also gives equally good agreement with the
inelastic spectrum, while correctly reproducing the
experimental ground state \cite{footnote1}.
The predicted inelastic scattering calculated from linear spin wave theory
for the parameters in Eq. (\ref{eq:Jparams}) is plotted in  Fig.~\ref{fig:inelastic}
for comparison with Ref. \cite{petit16}.

The theory presented here is also capable of accounting for
the positive 
Curie-Weiss temperature of Nd$_2$Zr$_2$O$_7$ \cite{bloete69, xu15, lhotel15, ciomaga15}, 
in spite of the antiferromagnetic ground state.
Specifically, the Curie-Weiss temperature for the model Eq. (\ref{eq:HXYZ}) is
\begin{eqnarray}
T_{\sf CW}=\frac{1}{2 k_B}\left(
\tilde{\sf J}_z \cos^2(\vartheta) + \tilde{\sf J}_x \sin^2(\vartheta)
\right).
\end{eqnarray}
where $k_B$ is Boltzmann's constant.
If we take $\vartheta\approx0.83$ this reproduces the Curie-Weiss
temperature $T_{\sf CW}\approx0.2$K measured in Ref. [\onlinecite{xu15, lhotel15, ciomaga15}].
As a final consistency check, we can then calculate the magnitude of the
ordered moment $m^{\sf ord}$ which we expect to find in the ground state.
At our level of approximation, the ratio of $m^{\sf ord}$ 
to the full, saturated, moment $m^{\sf sat}$  is given by
\begin{eqnarray}
\frac{m^{\sf ord}}{m^{\sf sat}}=\cos(\vartheta)\left(\frac{S-\langle a_i^{\dagger} a_i \rangle}{S} \right).
\end{eqnarray}
The spin wave calculation gives us $\left(\frac{S-\langle a_i^{\dagger} a_i \rangle}{S} \right) \approx 0.87$.
The fact that this number is close to unity is a good indicator that the linear spin wave 
approach is valid.
Combining this with the estimated value of $\vartheta$ gives an
ordered moment fraction $\frac{m^{\sf ord}}{m^{\sf sat}}\approx0.59$.
This is close to the value $\frac{m^{\sf ord}}{m^{\sf sat}}\approx0.5$ obtained in
Ref. [\onlinecite{xu15}], but a bit higher than the value  $\frac{m^{\sf ord}}{m^{\sf sat}}\approx0.33$ 
obtained in Ref. [\onlinecite{lhotel15}].
It is interesting to note that most of this moment reduction comes from the pseudospin
rotation $\vartheta$, not from the zero-point fluctuations, as one might typically expect.

The theory presented in this article is thus the first to
present a consistent treatment of the ground state and the finite energy spectrum
in Nd$_2$Zr$_2$O$_7$.
At the same time it is also able to account for the apparent contradiction between
the Curie-Weiss temperature and the antiferromagnetic ordering and 
gives reasonable agreement with the strongly reduced ordered moment measured
in experiments \cite{xu15}.

{\it Moment fragmentation}--
Having established that a theory based on a linear spin wave treatment
of Eq. (\ref{eq:HXYZ}) correctly reproduces the experimental
phenomenology, we can now ask what this theory tells about the proposed
``magnetic moment fragmentation''. 
In particular, is this a true example of moment fragmentation, as proposed
in Ref. \cite{brooks14}, and if so, what is its origin?

\begin{figure}
\centering
\includegraphics[width=0.75\columnwidth]{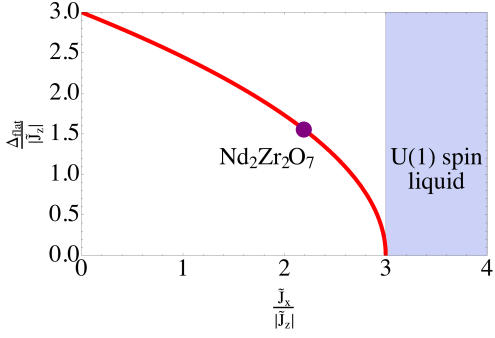}
\caption{Closing of the flat band gap $\Delta_{\sf flat}$ with increasing $\tilde{J}_x$ 
and the proximity of a $U(1)$ quantum spin liquid phase, shown for $\tilde{J}_y=0$. 
The gap to the flat band which contains the Coulomb-phase-like correlations of
Fig. \ref{fig:pinchpoints}(b) closes when the ratio $\frac{\tilde{J}_x}{|\tilde{J}_z|}=3$.
This is a likely indicator of the onset of a $U(1)$
spin liquid phase.
Based on the parameterisation of Eq. (\ref{eq:Jparams}) this ratio is 
$\frac{\tilde{J}_x}{|\tilde{J}_z|}\approx2.19$ in Nd$_2$Zr$_2$O$_7$, suggesting 
 the possibility of observing a $U(1)$ spin liquid ground state
Nd$_2$Zr$_2$O$_7$ or related Nd based pyrochlores,  induced by
application of chemical or physical pressure.
}
\label{fig:gap}
\end{figure}

The proposal of Brooks-Bartlett {\it et al.} in Ref. \cite{brooks14} is based on
the Helmholtz decomposition of the magnetisation density
\begin{eqnarray}
{\bf m}={\bf m}_m + {\bf m}_d=\nabla \psi+ \nabla \times {\bf A}
\label{eq:helmholtz}
\end{eqnarray}
where ${\bf m}_d= \nabla \times {\bf A}$ is divergence-free ($\nabla \cdot {\bf m}_d=0$) and 
${\bf m}_m=\nabla \psi$ is ``divergence-full''.
The fragmentation phenomenon is observed when magnetic order
occurs in ${\bf m}_m$, but ${\bf m}_d$ remains fluctuating quasi-independently
of   ${\bf m}_m$. 
Since ${\bf m}_d$ obeys $\nabla \cdot {\bf m}_d=0$
this gives rise to the pinch point correlations associated with a Coulomb
phase \cite{henley10, isakov04, henley05, fennell09}.

We can understand the magnetic fragmentation phenomenon in
Nd$_2$Zr$_2$O$_7$ by defining fields ${\bf m}_i^{(\tilde{\alpha})}$
for each pseudospin component  $\tilde{\tau}^{\tilde{\alpha}}_i$, according to Eq. (\ref{eq:xyz-mfields})
and then applying the Helmholtz decomposition to each one individually
\begin{eqnarray}
{\bf m}_i^{(\tilde{\alpha})}=
\nabla \psi^{({\tilde{\alpha}})} + \nabla \times {\bf A}^{({\tilde{\alpha}})}.
\label{eq:indy-helmholtz}
\end{eqnarray}
(Note that ${\bf m}_i^{(\tilde{y})}$ does not contribute to the
physical magnetisation field ${\bf m}_i$ [Eq. (\ref{eq:magnetisation-3})]).

In the all-in-all-out ground state, ${\bf m}_i^{(\tilde{z})}$
is completely ``divergence-full'', and we may write $ {\bf A}^{({\tilde{z}})}=0$.
The fluctuations of ${\bf m}_i^{(\tilde{x})}$ and ${\bf m}_i^{(\tilde{y})}$,
on the other hand, have both divergence-free
and ``divergence-full'' components.
The moment fragmentation phenomenon is observed because the
equations of motion decouple the dynamics of divergence-free and
``divergence-full'' components of ${\bf m}_i^{(\tilde{x})}, {\bf m}_i^{(\tilde{y})}$.

Writing down the Heisenberg  equations of motion for
${\bf m}_i^{(\tilde{x})}$ and
${\bf m}_i^{(\tilde{y})}$
and linearising around the all-in-all-out ground state we find
\begin{eqnarray}
&&
\partial_t {\bf m}_i^{(\tilde{\alpha})}
\approx \varepsilon_{\tilde{\alpha}'\tilde{\alpha} \tilde{z}} S \big( \tilde{{\sf J}}_{\alpha'}\nabla_i(\nabla \cdot {\bf m}^{(\tilde{\alpha}')}) +
 \nonumber \\
&& \qquad \qquad  \qquad\qquad\quad 
(6 |\tilde{\sf J}_z|-2\tilde{\sf J}_{\alpha'})
{\bf m}_i^{(\tilde{\alpha}')} \big)
\label{eq:EOM}
\end{eqnarray}
where $\tilde{\alpha}'=\tilde{y}$ when $\tilde{\alpha}=\tilde{x}$
and vice versa and $\varepsilon_{\tilde{x}\tilde{y} \tilde{z}}=-\varepsilon_{\tilde{y}\tilde{x} \tilde{z}}=1$.
In Eq. (\ref{eq:EOM}), $\nabla$ and $\nabla \cdot$ should be
interpreted as the lattice gradient and divergence.
This suggestive form for the equations of motion in terms of the 
 lattice gradient and divergence
arises because
the sites of the pyrochlore lattice can be considered as the bonds of
a bipartite (in this case, diamond) lattice \cite{henley10}.

Eq. (\ref{eq:EOM}) can be solved in terms of the Helmholtz decompositions
[Eq. (\ref{eq:indy-helmholtz})], by writing
\begin{eqnarray}
&&\partial_t \psi^{(\tilde{\alpha})}=\varepsilon_{\tilde{\alpha}'\tilde{\alpha}\tilde{z}}  S\big(\tilde{{\sf J}}_{\alpha'} 
\nabla^2 \psi^{(\tilde{\alpha}')}+  
(6 |\tilde{\sf J}_z|-2\tilde{\sf J}_{\alpha'})
\psi^{(\tilde{\alpha}')} \big) \nonumber \\
\label{eq:helmholtzmotionPsi} \\
&& \partial_t {\bf A}^{(\tilde{\alpha})}=\varepsilon_{\tilde{\alpha}'\tilde{\alpha} \tilde{z}} S
(6 |\tilde{\sf J}_z|-2\tilde{\sf J}_{\alpha'})
{\bf A}^{(\tilde{\alpha}')}
\label{eq:helmholtzmotionA}
\end{eqnarray}

The important feature of Eqs. (\ref{eq:helmholtzmotionPsi})-(\ref{eq:helmholtzmotionA}) is that the divergenceless
fluctuations \big(i.e. fluctuations of  $ {\bf A}^{(\tilde{\alpha})}$\big)
are completely decoupled from the ``divergence full'' fluctuations \big(i.e.
fluctuations of $\psi^{(\tilde{\alpha})}$\big).
Fluctuations of $ {\bf A}^{(\tilde{\alpha})}$ form a flat band
at energy $\Delta_{\sf flat}=\sqrt{(3 |\tilde{\sf J}_{ z}|-\tilde{\sf J}_{ x})(3 |\tilde{\sf J}_{ z}|-\tilde{\sf J}_{ y})}$, while fluctuations of $\psi^{(\tilde{\alpha})}$ form
dispersive bands.

The physical magnetisation field in Nd$_2$Zr$_2$O$_7$ [Eq. (\ref{eq:magnetisation-3})]
 thus comprises (i) a static, ordered,
``divergence full'' component, 
(ii) a finite energy divergenceless component exhibiting
Coulomb-liquid-like correlations and finally
(iii)
another ``divergence full'' component corresponding to the finite energy dispersive bands.

The fact that all three components are observable within a magnetisation field
which is strictly Ising like (in the sense that ${\bf m}_i$ is always parallel to the local easy 
axis) is a consequence of the unusual symmetry of dipolar octupolar doublets- specifically that
the $x$-component of the pseudospin transforms like the $z$-component of a dipole moment.
This understanding of the moment fragmentation is fully compatible with the observation
that the pinch points remain observable above the ordering transition at $T_N$, but at
lower frequency \cite{petit16}.
Above the transition ${\bf m}_i^{(\tilde{x})}$ can fluctuate for little or no
energy cost but its correlations will remain ice like due to the positive value of $\tilde{\sf J}_x$.

{\it Conclusions}-- In conclusion we have explained the quantum origins of
the moment fragmentation in Nd$_2$Zr$_2$O$_7$, observed in Ref. \cite{petit16}.
It may be rationalized as the consequence of the symmetry properties of dipolar-octupolar
doublets and a decoupling of divergence-free and divergence-full fluctuations in the
equations of motion.

Much of the physics discussed here is generic to systems described by
the exchange Hamiltonian $\mathcal{H}_{\sf XYZ}^{\sf DO}$ which have an all-in-all-out
ground state. 
Specifically, the flat band exhibiting pinch points at finite energy is present throughout the
all-in-all-out phase of $\mathcal{H}_{\sf XYZ}^{\sf DO}$, at least at the level of linear spin wave theory.
It will therefore be interesting to investigate whether the moment fragmentation phenomenon
is also observed in other Nd based pyrochlores showing an all-in-all-out ground state such as 
Nd$_2$Sn$_2$O$_7$ \cite{bertin15}, Nd$_2$Hf$_2$O$_7$ \cite{anand15}
and possibily Nd$_2$Pb$_2$O$_7$ \cite{hallas15}.

The parameterisation of the exchange Hamiltonian $\mathcal{H}_{\sf XYZ}^{\sf DO}$ 
[Eq. (\ref{eq:HXYZ})], given in Eq.~(\ref{eq:Jparams}) suggests that Nd$_2$Zr$_2$O$_7$ is
proximate to the $U(1)$ spin liquid phase which has been long sought amongst
``quantum spin ice'' pyrochlores
\cite{gingras14, savary12, hermele04, banerjee08, benton12, shannon12, hao14, kato15}.
As shown in Fig. \ref{fig:gap},
the closing of the gap to the flat band containing the pinch point correlations occurs 
at $\frac{\tilde{\sf J}_x}{|\tilde{\sf J}_z|}=3$ within linear spin wave theory.
Classically, this would signal the formation of
an extensive ground state manifold with ice-like
character, but the
mixing of these states by quantum fluctuations is known to stablise a $U(1)$ spin liquid
with dynamic emergent gauge fields \cite{hermele04, shannon12}.
The placement of  Nd$_2$Zr$_2$O$_7$ close to the point where this gap vanishes 
hints at the proximity of the $U(1)$ spin liquid phase.
If there is a well formed Coulomb phase above $T_N$ in Nd$_2$Zr$_2$O$_7$ this may make 
the observed magnetic ordering a candidate for the observation of a Higgs transition in which
the emergent gauge field of the Coulomb phase is gapped by the condensation of emergent
gauge charges \cite{powell11, chang12}.
We therefore have reason to hope that experiments on Nd$_2$Zr$_2$O$_7$ and related materials
may yet reveal even more exotic phenomena.


{\it Acknowledgements} --
This work was supported by the Theory of Quantum Matter Unit of the 
Okinawa Institute of Science and Technology Graduate University.
The author is grateful to Ludovic Jaubert and Nic Shannon for careful readings
of the manuscript.




\begin{thebibliography}{10}
\bibitem{footnote1}
The two parameterisations can be related by a unitary transformation
$\tau_{\alpha} \to U^{\dagger}\tau_{\alpha}U$
where the unitary operator $U$ corresponds
to a $\frac{4\pi}{3}$ rotation
around the $(1, 1, 1)$ direction of pseudospin space (note that this
not the same as the $(1, 1, 1)$ direction of crystal coordinate space).
\end{thebibliography}
\end{document}